\documentclass[10pt,letterpaper,twocolumn]{article} 

\usepackage{ol}
\usepackage{hyperref}
\usepackage{amsmath}
\usepackage{epsfig}

\begin{document}

\twocolumn[

\title{Time-resolved refractive index and absorption mapping of light-plasma filaments in water}

\author{Stefano Minardi, Amrutha Gopal, Michael Tatarakis}

\address{Department of Electronics, Technological Educational Institute of Crete - Romanou, 3 -
GR73133 Chania, Greece}

\author{Arnaud Couairon}

\address{Centre de Physique Th{\'e}orique, CNRS, {\'E}cole Polytechnique, F-91128 Palaiseau, France}

\author{Gintaras Tamo\v{s}auskas, Rimtautas Piskarskas, Audrius Dubietis, and Paolo Di Trapani}

\address{Department of Quantum Electronics, Vilnius University - Sauletekio, 9 - bldg. 3, LT-10222 Vilnius, Lithuania}


\begin{abstract} By means of a quantitative shadowgraphic method, we performed a space-time characterization of the refractive index variation and transient absorption induced by a light-plasma filament generated by a 100 fs laser pulse in water. The formation and evolution of the plasma channel in the proximity of the nonlinear focus were observed with a 23 fs time resolution.
\end{abstract} 


]

Structural modifications or changes of the index of refraction induced by focusing an ultrashort laser pulse in a transparent dielectric is the basis of modern production of optical devices buried in the bulk of optical media, such as waveguides \cite{DavisOL96}, gratings\cite{SudrieOC99}, three dimensional optical memories\cite{GlezerOL96}, as well as of biomedical applications such as refractive surgery \cite{ArnoldAPB2005}.
Around the focus, the laser pulse usually generates pairs of electron-holes by optical field ionization. For tight focusing configuration, the electron density is in turn all the more enhanced by inverse Bremsstrahlung reaching easily $10^{20}$ cm$^{-3}$ and inducing strong absorption.\cite{CouaironPRB05,TzortzakisPRL}
For collimated beams or small numerical apertures, the electron hole plasma may be limited to densities of a few $10^{18}$ cm$^{-3}$ (if excited at $\lambda=800$ nm), thereby allowing weak absorption and filamentary propagation of the laser beam. Post mortem of damaged materials provides an indirect estimation of the electron plasma density but its direct measurement is a difficult task in the filamentary regime. \cite{Couairon06}
The small transverse size of the plasma channel ($\simeq 2-5\mu$m) and the correspondingly small induced phase shift ($\pm 0.01$ radians) make it difficult to perform interferometric or all-optical spatio-temporally resolved measurements. Previous measurements \cite{TzortzakisPRL,Sun05} addressed only the tight-focusing configuration in solids which leads to a large size (tens of microns) and high density plasma distribution ($\simeq 10^{19}$ cm$^{-3}$). Mao {\it et al.} proposed to characterize plasma densities by absorptive shadowgraphy \cite{MaoMao}, but the method requires a reliable calibration sample to retrieve quantitative information.    
Yet, precise measurement of the electron density generated by loosely focusing an ultrashort laser pulse in a dielectric is required for applications where non-permanent changes in the refractive index must be performed by single shot longitudinal illumination of optical material   without damage. As for fundamental research, the role of a tenuous plasma channel in the dynamics of filamentation might be less significant in condensed media than in gases, as suggested by the recent reinterpretation of this phenomena on the basis of the X-wave paradigm \cite{KolesikPRL04,FaccioPRE05,FaccioPRL06}. 

In this Letter, we present the first spatio-temporally resolved study of refractive index changes induced by a light-plasma filament in water, which we use as a convenient prototype for condensed media. The measurements were performed by means of a time-resolved quantitative shadowgraphic method sensitive to gradients of refractive index\cite{Hutchinson87} and recently used to retrieve high resolution electron density distribution of laser-induced plasma in air \cite{Gopal07}.
The technique is extremely sensitive in detecting the tiny refractive index changes induced by the optical Kerr effect and the plasma ($\Delta n\approx 10^{-4}-10^{-3}$) provided that the object is small enough and diffraction effects are negligible\cite{Trainoff02}. 
 
\begin{figure}[tb]
\centerline{\includegraphics[width=8.3cm]{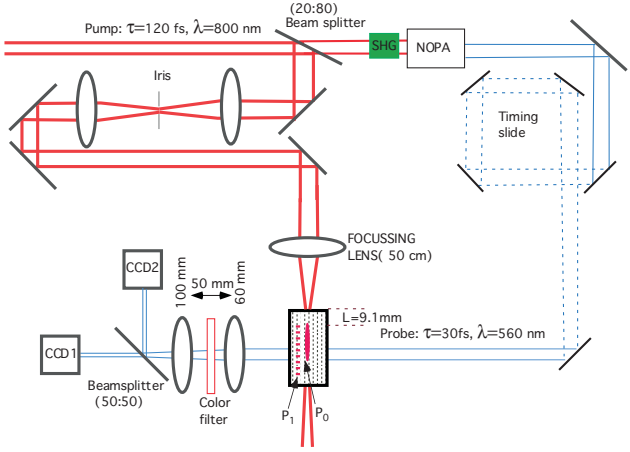}}
 \caption{\label{setup} Experimental set-up.}
\end{figure}

The experiment was carried out with an amplified Ti:Sapphire laser system (Spectra Physics) delivering $\lambda=800$ nm, $100$ fs pulses at 1 kHz rate. Our experimental set-up is shown in Fig.\ref{setup}. 
A spatially filtered laser pulse was loosely focused (beam waist diameter $d_{FWHM}=109 \mu$m) on the front surface of a 20 mm long fused silica cuvette filled with de-ionized water so as to generate
a filament of several millimeters.  For a pulse energy of $4\pm0.25 \mu$J, the nonlinear focus was located at 9.10 mm beyond the input window of the cuvette. The changes in the refractive index induced by the filament were probed transversally
by $\lambda=560$ nm, 23-fs-long (FWHM), spatially filtered pulses from a non-collinear optical parametric amplifier (TOPAS-White - Light Conversion Ltd.) pumped by the second-harmonic of the laser. The probe pulses were delayed by a motorized time slide and negatively pre-chirped in order to compensate for the dispersion due to the cuvette window and 2.5 mm of propagation in water. 

The maps of the refractive index variations induced by the filament were obtained by modifying the method outlined in Ref. \cite{Gopal07} so as to simultaneously image two planes, $P_0$ and $P_1$ in order to avoid systematic errors due to the absorption of the sample and reduce the impact of shot-to-shot fluctuations of the back-illuminating beam.
To record the shadowgraphs, we used a high spatial-resolution imaging set-up based on two $10$ bit CCD cameras. The plane $P_0$ containing the filament was imaged on the first camera whereas the parallel plane $P_1$  located $\simeq 100 \mu$m away from $P_0$ was imaged on the second camera (see Fig. \ref{setup}). The slight difference in the magnification factors of both channels is negligible ($<1\%$).
Both cameras were synchronized electronically with the laser trigger and could grab single shot images of the filament. 
The resolving power of the optical system was estimated experimentally to be $\simeq 1.5 \mu$m. 
The refractive index maps are finally obtained by Abel inversion\cite{Hutchinson87} of the phase data retrieved from the shadowgrams.

\begin{figure}[tb]
\centerline{\includegraphics[width=8.3cm]{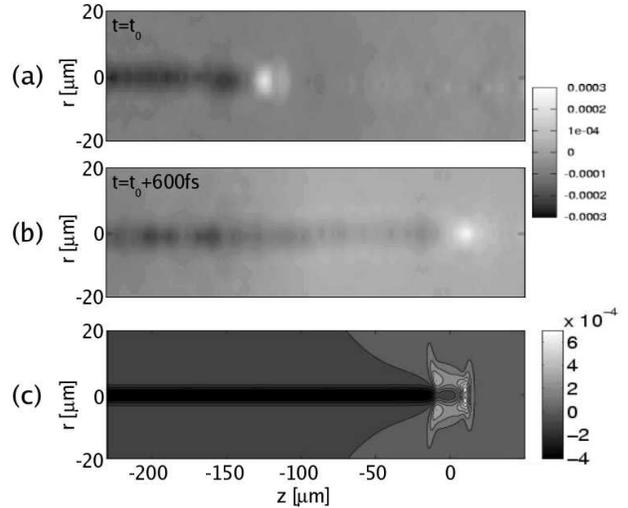}}
 \caption{\label{2Dmap} (a) and (b) Refractive index map of the filament taken at two different times and showing the formation of the plasma channel. Horizontal scale: propagation axis ranging from 9.20 mm (left) to 9.48 mm (right) from the input window. (c) Refractive index map obtained by numerical simulation of the experiment.}
\end{figure}

Figures \ref{2Dmap} (a) and (b) show a typical measurement of the filament track relative to a position a few hundreds of microns beyond the nonlinear focus. 
The head of the filament is marked by a peak ($5 \mu$m $\times$ $40$ fs, FWHM) with positive variation of the refractive index (peak value $\Delta n \simeq +3.0\times 10^{-4}$), followed by a narrow channel (average FWHM diameter $5.5\pm1.5 \mu$m) with $\Delta n \simeq (-1.2\pm0.2)\times 10^{-4}$.
We attribute the positive index change as due to the optical Kerr effect,
while the negative region is a clear signature of the existence of an electron plasma.
Assuming the refractive index of the plasma to follow the Drude model\cite{Hutchinson87}, the observation is consistent with an electron density of about $n_e\simeq (2.1\pm0.4) \times 10^{18}$ cm$^{-3}$. The depth and width of the plasma channel exhibits a modulation on the $\approx 100 \mu$m scale, which is attributed to the fine-structure of the plasma channel \cite{Dubietis06}.  
The speed of the refractive index peak measured from our measurements at different probe delays was found to be $0.2235\pm0.0005$ $\mu$m/fs or $(0.7455\pm 0.0016)c$, in agreement with the expected group velocity of light in water.

We compared our experimental data with a numerical simulation. The filamentation of the laser pulse was modeled by means of a nonlinear propagation equation for the pulse envelope which accounts for diffraction and dispersion, Kerr nonlinearity ($n_2=1.6\times 10^{-16}$ cm$^2$/W), self-steepening, multiphoton absorption (MPA; order: $K=5$; cross-section: $\beta_K=8.3 \times 10^{-50}$ cm$^7$/W$^4$) and plasma generation via MPA and inverse Bremsstrahlung (collision time: $\tau_c=3$ fs); see Refs. \cite{Dubietis06,Couairon06} for details. 
Fig. 2(c) shows the refractive index map obtained by simulation for input conditions as in the experiment. The set of parameters used in the model  allowed us to reproduce with very good quantitative agreement the features of the filament (nonlinear focus, length, diameter) visible on the map in Fig. 2. The computed peak electron density was $3\times 10^{18}$ cm$^{-3}$; the measurement technique therefore provides reliable information on the density of rather tenuous plasmas generated in condensed media.

\begin{figure}[tb]
\centerline{\includegraphics[width=8cm]{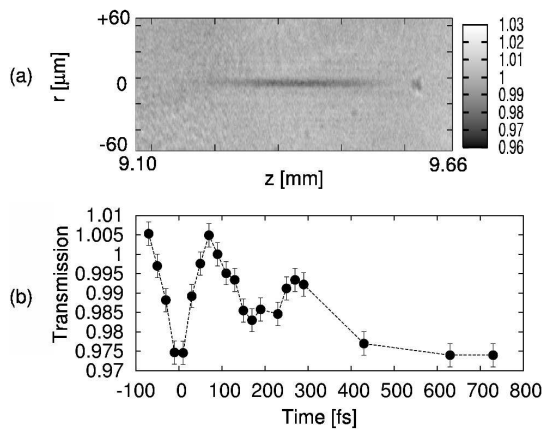}}
 \caption{Absorption measurements of the filament at 560 nm. (a) As a function of position for a given probing time; (b) as a function of time at a given position (z=9.54 mm). Circles: experimental data. The line is a guide to the eye.}
\end{figure}

An inspection of the raw shadowgrams taken at the plane $P_0$ provides useful information on the absorption of the probe pulse in the filament (Fig. 3(a)). 
In particular, we observe three features: \textit{i)} a strongly absorbing and localized head (corresponding to the peak of positive $\Delta n$) followed by \textit{ii)} a zone with almost no absorption and \textit{iii)} a tail with slowly varying absorption (up to 3.0\%). Figure 3(b) shows the on-axis transmittance of the probe pulse as a function of time, at a given propagation distance from the input (z=9.54 mm). 
  
We interpret the low transmittance of the head of the filament as the consequence of the simultaneous absorption of one photon of the probe ($\lambda_1 = 560 $ nm) and three photons of the pump beam ($\lambda_0 = 800 $ nm). Indeed, this MPA transition is already above the gap for indirect ionization of water ($E_{gap}=$6.5 eV)\cite{Lian05}. 

For what concerns the plasma channel in the wake of the pulse, we note that our observations cannot be explained relying only on the
mechanism of inverse Bremsstrahlung, as often assumed.
In fact, for the measured plasma density and under the
assumption of electron collision time of 3 fs\cite{Dubietis06}, inverse
Bremsstrahlung would lead to a maximum absorption
of 0.2\%. Moreover, the process does not justify the long
build-up time of the absorption ($\simeq 0.5-0.7$ ps) and its
very slow decay (the absorption in the tail of the filament lasted for delays up to 90 ps after the passage of the optical pulse). As a possible explanation we propose to consider the role of the transient absorption induced by solvated electrons\cite{Lian05,Elles06}.

Due to the high polarity of water, a considerable fraction (up to 20-30\%) of the electrons formed by MPA should be solvated, \textit{i.e.} trapped in a cage of water molecules oriented to screen the free electron charge. 
The solvated states are reported to be long lived ($t_{decay}> 500$ ps), and are characterized by an optical resonance centered at $\lambda = 720$ nm which builds up in about 1.0-2.0 ps from the ionization event \cite{Lian05}, leading to an absorption dynamic consistent with our observations.  

In conclusion, we have shown that quantitative shadowgraphy is a powerful tool for time resolved refractometry of laser-plasma filaments in condensed media. This technique enabled us for the first time to take snapshots of the propagation of a fs pulse in water and trace the dynamic of the formation of the plasma channel with a time resolution better than the pump pulse duration. 
Useful quantitative information such as the size and electron density of the plasma channel generated by the writing pulse were deduced from the measured refractive index change and compared to results of numerical simulations of ultrashort pulse filamentation. 
Moreover, spatially resolved transient absorption measurements bear evidence of solvation effects of the electrons released during light-water interaction, providing useful insight in the chemical changes induced by femtosecond laser pulse propagation in water.    

The authors would like to acknowledge the support from the Access to Research Infrastructures activity in the 6th Framework Programme of the EU  (Contract RII3-CT-2003-506350, VULRC - Laserlab Europe), from the Marie Curie Transfer of Knowledge grant DAIX Contract MTKD-CT-2004-014423, and the Marie Curie Chair project STELLA, Contract MEXC-CT-2005-025710 (www.vino-stella.eu). A special thank also to Dr. Gediminas Veitas for helping in setting up the experiment, and to Dr. Gloria Tabacchi and Dr. Matteo Cavalleri for stimulating discussions.

\twocolumn[

]
\end{document}